# CNN-based forecasting of early winter NAO using sea surface temperature

Elena Provenzano[1], Guillaume Gastineau[1], Carlos Mejia[1], Didier Swingedouw[2] and Sylvie Thiria[1]

[1]UMR LOCEAN, IPSL, Sorbonne Université, IRD, CNRS, MNHN, Paris, France
[2]Environnements et Paléoenvironnements Océaniques et Continentaux (EPOC), Université de Bordeaux, Pessac, France
**Corresponding author:** Elena Provenzano; Email: elena.provenzano@locean.ipsl.fr

## Abstract

The North Atlantic Oscillation (NAO) is the dominant mode of atmospheric variability over the North Atlantic, influencing temperature and precipitation across Europe. While the NAO's impact on North Atlantic sea surface temperatures (SSTs) is well understood, the NAO can also be driven by SST anomalies, although this response is believed to be weak and nonlinear. Former studies highlight that during early winter (November–December), El Niño-Southern Oscillation (ENSO) events modulate the NAO, with El Niño (La Niña) events being linked to positive (negative) NAO phases, and an opposite effect in late winter (January–February). Indian Ocean SSTs and the North Atlantic Horseshoe SST anomaly have also been suggested as contributors to early winter NAO variability. However, climate models often struggle to capture these SST-NAO teleconnections, particularly in early winter. To address this, a statistical framework based on convolutional neural networks (CNNs) is developed to predict the early winter NAO using observed SST fields one, two, and three months before. A linear model serves as a benchmark, and both are trained on ERA5 reanalysis data from 1940 to 2023. A sensitivity analysis interprets the CNN's decision-making process, revealing that it focuses on regions such as the tropical Pacific and North Atlantic, confirming earlier findings. The CNN outperforms the linear model, highlighting the value of capturing nonlinear SST-NAO relationships. Prediction skill appears to be linked to ENSO, with strong events associated with greater skill than neutral events. These findings underscore the potential of deep learning to build NAO prediction.

## Impact Statement

The North Atlantic Oscillation (NAO) drives a large part of the seasonal climate variability across Europe, North Africa, and North America, with significant impacts on sectors such as energy and agriculture. Sea surface temperatures (SST) are suggested to influence the NAO through atmospheric teleconnections, but this relationship is nonlinear, not fully understood, and not well represented in forecast systems. To improve our understanding of the role of SST in NAO predictability, this work employs convolutional neural networks that only use observed SST to predict the NAO. This approach highlights the importance of accounting for nonlinear interactions, particularly those

This research article was awarded Open Data and Open Materials badges for transparent practices. See the Data Availability Statement for details.

influenced by the asymmetric and nonlinear nature of the ENSO phenomenon. The research contributes to a deeper understanding of ocean–atmosphere dynamics, with potential applications for the use of data-driven approaches in climate forecasting across time scales ranging from 10 days to three months.

## 1. Introduction

The North Atlantic Oscillation (NAO) is the dominant mode of atmospheric variability over the North Atlantic for time scales larger than 10 days (e.g., Barnston and Livezey, 1987; Feldstein, 2000), characterized by a dipole pattern of sea-level pressure anomalies that significantly influence temperature and precipitation anomalies across Europe (Scaife et al., 2008). The NAO owes its existence to internal atmospheric dynamics but can be influenced by several external drivers. Among them, sea surface temperature (SST) can affect the NAO, although this effect is weak and nonlinear (Kushnir et al., 2002).

During early winter, it is suggested that the NAO responds to a tripolar SST anomaly, commonly referred to as the North Atlantic Horseshoe (Gastineau and Frankignoul, 2015). Remote SSTs also play a role, most notably from the El Niño-Southern Oscillation (ENSO), the dominant mode of interannual climate variability at the global scale. ENSO affects global circulation patterns, and in early winter (November–December), El Niño (La Niña) events have been linked to positive (negative) NAO phases, with an opposite relationship observed in late winter (January–February) (King et al., 2018; Zhang and Jiang, 2023). SST anomalies in the Indian Ocean have likewise been identified as significant contributors to early winter NAO variability (Hardiman et al., 2020), with studies showing that the Indian Ocean Dipole (IOD) acts as an independent source of forcing rather than merely a response to ENSO (Abid et al., 2023; Raganato et al., 2025).

Traditional climate models often struggle to capture the NAO's response to SSTs, particularly during early winter (Molteni and Brookshaw, 2023). To address this challenge, this study adopts a statistical modelling approach, employing convolutional neural networks (CNNs) to predict the NAO in early winter using leading observed SST fields as predictors. The objective is specifically aiming at assessing SST-driven predictability of the NAO, and not of other modes of atmospheric variability or associated atmospheric fields. This approach seeks to explore the potential of deep learning to predict NAO from preceding SST, and improve our understanding of ocean–atmosphere interactions, including the role of ENSO in modulating NAO predictability. CNN performance is compared against a benchmark linear model to assess whether the nonlinear approach provides improved skill.

## 2. Data and methods

### *2.1. Data and preprocessing*

Sea-level pressure (SLP) data from the ERA5 Hourly Data reanalysis (Hersbach et al., 2023) covering the period 1940–2023 is used to characterize the NAO. The spatial domain is limited to 20°N-70°N and 100°W-40°E. The original 0.25° × 0.25° resolution data is re-gridded to a coarser 2° × 2° grid using bilinear interpolation. Re-gridding to a coarser resolution is appropriate given the focus on large-scale variability and reduces the dimensionality of the dataset, thereby limiting potential overfitting associated with the relatively small sample size. Early winter (November–December) anomalies are computed by subtracting daily climatologies from individual days, and a 10-day centered running mean is then applied to these anomalies to remove high-frequency variability. SSTs from the same dataset are obtained for the corresponding temporal range within the 15°S-70°N spatial domain. SST is kept at a finer 1° × 1° resolution than SLP, as smaller-scale structures such as ocean fronts and eddies may provide additional predictive information for the NAO (Small et al., 2007). The anomalies are computed following the same procedure as for the SLP, except that the running mean uses a centered 29-day window instead of a 10-day window. The SST smoothing is applied to reduce the noise of the data while keeping the fluctuations at monthly to interannual scales, which are the focus of the present study.

The NAO index is defined as the standardized first principal component (PC1) from the Empirical Orthogonal Function (EOF) analysis of the SLP anomalies obtained from November–December, explaining 22.4% of the variance. Figure 1 shows the associated spatial pattern (EOF1), defined as a regression on the standardized PC1.

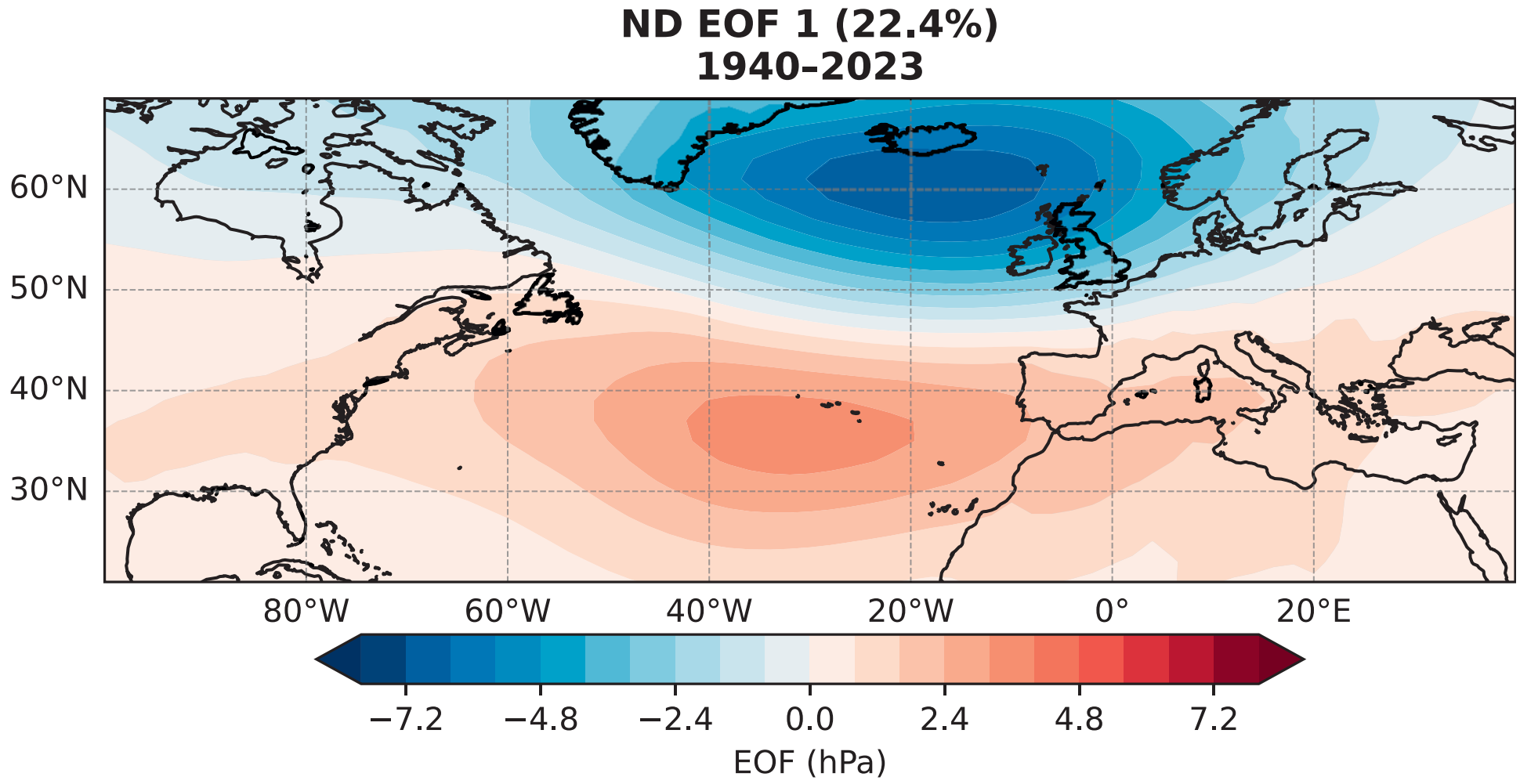

***Figure 1.*** *First EOF of North Atlantic SLP anomalies (in hPa), computed from daily 10-day running mean time series for early winter (November–December) in 1940–2023. The variance explained is given in parentheses at the top.*

### *2.2. Convolutional neural network (CNN)*

The data is used to train a CNN that takes three SST fields as input, corresponding to one, two, and three months before the predicted NAO index, in order to estimate it. This setup is designed to capture the atmospheric response to the ocean rather than the dominant influence of the NAO in driving the SST anomalies.

Each SST field is preprocessed with a centered 29-day running mean filter, which smooths daily SST values by averaging over a ± 15-day window. The target NAO index is derived from SLP anomalies smoothed with a centered 10-day running mean. As a result of both centered running means, the SST input at month t-1 contains information up to 15 days after the nominal lag, while the NAO target is centered over a ± 5-day window, resulting in an effective temporal offset of 10 to 110 days between input and output.

While the temporal offsets introduced by the filtering approach overlap with the medium-range forecasting window, our aim here is not to produce operational forecasts. Rather, we seek to identify the SST signals that precede atmospheric circulation variability in the Atlantic on sub-seasonal to seasonal time scales.

The use of three SST input fields (the one-, two-, and three-month lags), rather than one or two, improves prediction skill (not shown). However, larger lags are not considered, as the typical time scale of the atmospheric response to SST ranges from a few days to approximately two months (Deser et al., 2007; see review of Yuan et al., 2018). The SST data is already expressed as anomalies with typical magnitudes of the order of 1 °C, while the NAO index is computed from the standardized PC1, so that no additional preprocessing is required to normalize the data. The CNN model employed in this study consists of three convolutional blocks, each incorporating circular padding along the longitude dimension, batch normalization, and the ReLU activation function. The convolutional layers use kernel sizes of 5 × 3, 3 × 3, and 3 × 3, respectively.

Through a series of down-sampling operations, the spatial dimensions are progressively reduced from the original input size of 85 × 360 (latitude × longitude) while preserving key features. The output of the final convolutional block, with dimensions 17 × 59, is flattened and passed to a fully connected regression head with a linear activation that produces the model's final output, i.e., a single real number representing the NAO index (Figure 2). In total, the model contains 262,225 trainable parameters.

A linear equivalent of the CNN is constructed as a benchmark for comparison by removing all nonlinear activation functions from the architecture. All ReLU activations are eliminated and replaced with Identity operations so that each layer performs only linear transformations. Max pooling is also nonlinear because it selects the maximum value within each region, so it was replaced with average pooling, which computes the mean and therefore preserves linearity. This ensures that the entire network behaves as a purely linear model. The comparison between these two models allows for a clear isolation of the added value of nonlinearities in the CNN. Additionally, the proposed linear model enables the estimation of skill from a considerably more complex linear framework than is typically employed in climate science.

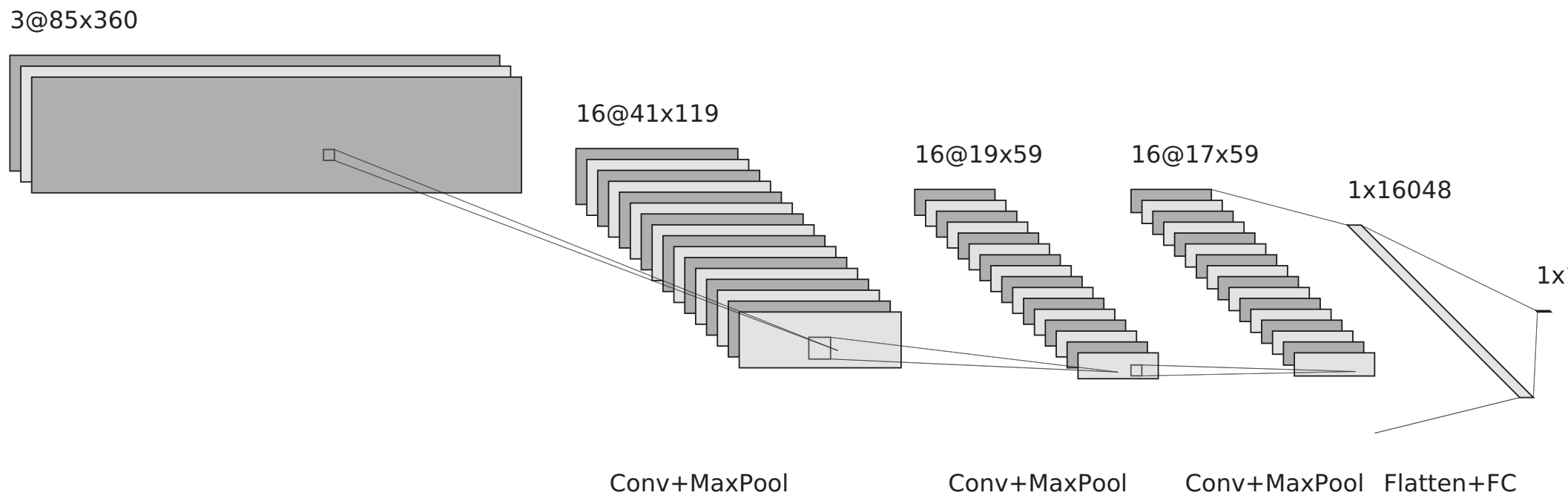

***Figure 2.*** *Diagram representing the CNN architecture, with the number of features and image dimensions (number of channels @ number of latitudes × number of longitudes) indicated above.*

### 2.3. Training and validation

The model is trained and validated using leave-one-out cross-validation (LOOCV), where each year from 1940 to 2023 is iteratively used as the validation set while all remaining years serve as the training set. This ensures that all years are included in validation, so that the resulting skill estimates are not dependent on the choice of validation years. This approach is especially well-suited for our data-limited case, as it maximizes the use of available data and provides a robust estimate of the model's generalization capability. The effective number of independent samples is estimated following Bretherton et al. (1999) as $N_{eff} = N/\tau$, where $\tau = 1 + 2\Sigma\rho(k)$ is the integral autocorrelation time scale of the SST summed to its first zero-crossing (lag = 41 days), giving $\tau = 43$ days and $N_{eff} = 118$ (about one to two independent samples per November–December season across 84 years). Training utilizes a batch size of 64, a learning rate of $2 \times 10^{-6}$, and the Adam optimizer. The mean squared error (MSE) loss function is used, as standard practice for regression problems. Early stopping based on validation loss with a patience of 5 epochs is used. It should be noted that, since the early stopping procedure allows the validation year to influence checkpoint selection, the resulting metrics represent upper-bound estimates of performance. The LOOCV framework is therefore intended to explore model generalizability and the factors influencing performance across years, rather than to provide a fully unbiased estimate of predictive skill.

Hyperparameters are tuned using the first ten years (1940–1949), which are excluded from the computation of skill metrics to ensure independent evaluation.

Model training and validation are performed on daily values, giving 5124 training samples in total (84 years × 61 days/year); as noted above, these daily samples are not independent. To highlight low-frequency variability, the results are evaluated using monthly averages in the following. Specifically, early-winter predictions from each year's leave-one-out model are concatenated, and the correlations shown later are computed from the combined monthly values.

### 2.4. Sensitivity analysis

Gradient-based saliency mapping is used to perform a sensitivity analysis and generate visual explanations for the CNN model. This approach is a model-aware sensitivity method, meaning it leverages the internal structure and gradients of the neural network to quantify how small changes in each input feature affect the model's output (Baehrens et al., 2009; Bommer et al., 2024). In practice, the gradients of the predicted NAO index with respect to each input SST grid point are computed. The absolute gradients are then multiplied, grid point by grid point, by the absolute input values to produce attribution maps at full input resolution (85 × 360 grid points, 1° spatial resolution). Each grid-point value thus represents the model's local sensitivity weighted by the input anomaly, that is its contribution to the predicted NAO index.

Finally, the maps are averaged across multiple validation years to identify the regions that consistently influence the model's NAO predictions. Intensities are standardized prior to inter-model comparisons. A one-sample, one-tailed t-test is applied at each grid point to assess whether its mean attribution significantly exceeds a threshold, defined as the 90th percentile of the time-averaged attribution map for that channel. Significance is determined at the $p < 0.05$ level, with correction for multiple comparisons using the false discovery rate procedure (Wilks, 2006), which controls the expected proportion of false positives among rejected hypotheses.

## 3. Results

### *3.1 CNN and linear model comparison*

Model performance is evaluated using the normalized RMSE (nRMSE) computed by dividing RMSE by the standard deviation of the observed monthly NAO index to express model errors relative to the observed variability, and the Pearson correlation coefficient, both computed on the monthly aggregated NAO index time series. When evaluated over the entire validation period (i.e., 1950–2023), the CNN prediction system based solely on SST shows predictive skill for the NAO. It achieves an nRMSE of 0.88 and a Pearson correlation of 0.50, significant at the 95% confidence level, despite year-to-year variability in performance and a smaller variance compared to the observed values (Figure 3). Conversely, the baseline linear model records a higher nRMSE of 0.92 and a lower correlation of 0.44, also statistically significant.

Note that the linear model better predicts NAO values with a larger amplitude than the CNN, possibly because ReLU activations set negative values to zero and max pooling retains only local maxima, both of which can compress the amplitude of predicted extreme events. Specifically, it is of interest to notice that the strong positive NAO following the strong positive IOD event in 2019/20 (Hardiman et al., 2020) is not well predicted by either model. A hypothesis to explain this potential failure is that such strong IOD events may have been underrepresented in the training data, limiting the CNN's ability to capture their influence on the NAO during this particular season. Similarly, both models fail to accurately predict the NAO response during strong El Niño events such as 1997/98, likely due to the scarcity of such extreme events in the training data.

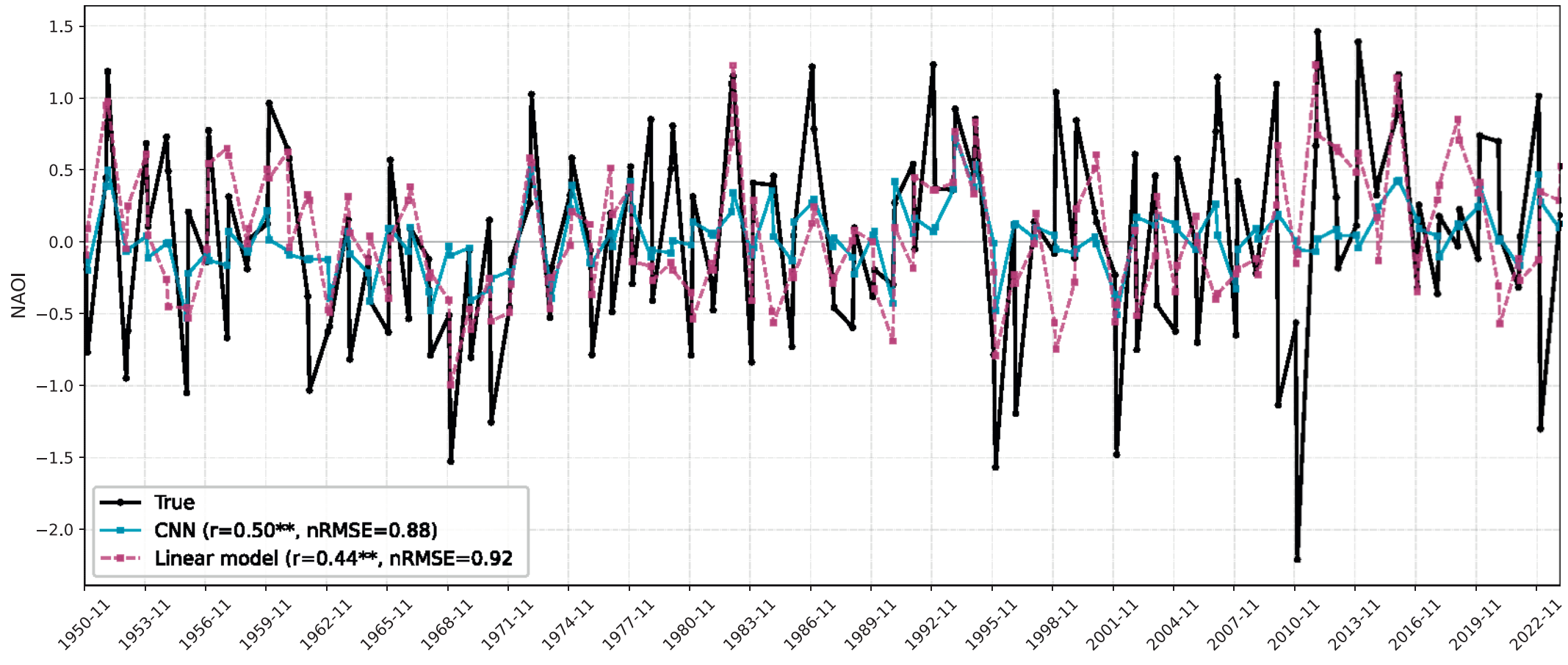

***Figure 3.*** *Time series of the monthly mean early winter (November–December) NAO index calculated from the daily estimates produced by the statistical models. Observed values (black) are compared to the convolutional neural network (CNN) predictions (light blue) and the linear model predictions (dashed purple line), obtained using leave-one-out cross-validation. The legend indicates the normalized RMSE (nRMSE) and correlation coefficient (r), and asterisks (**) denote significance at the 95% confidence level.*

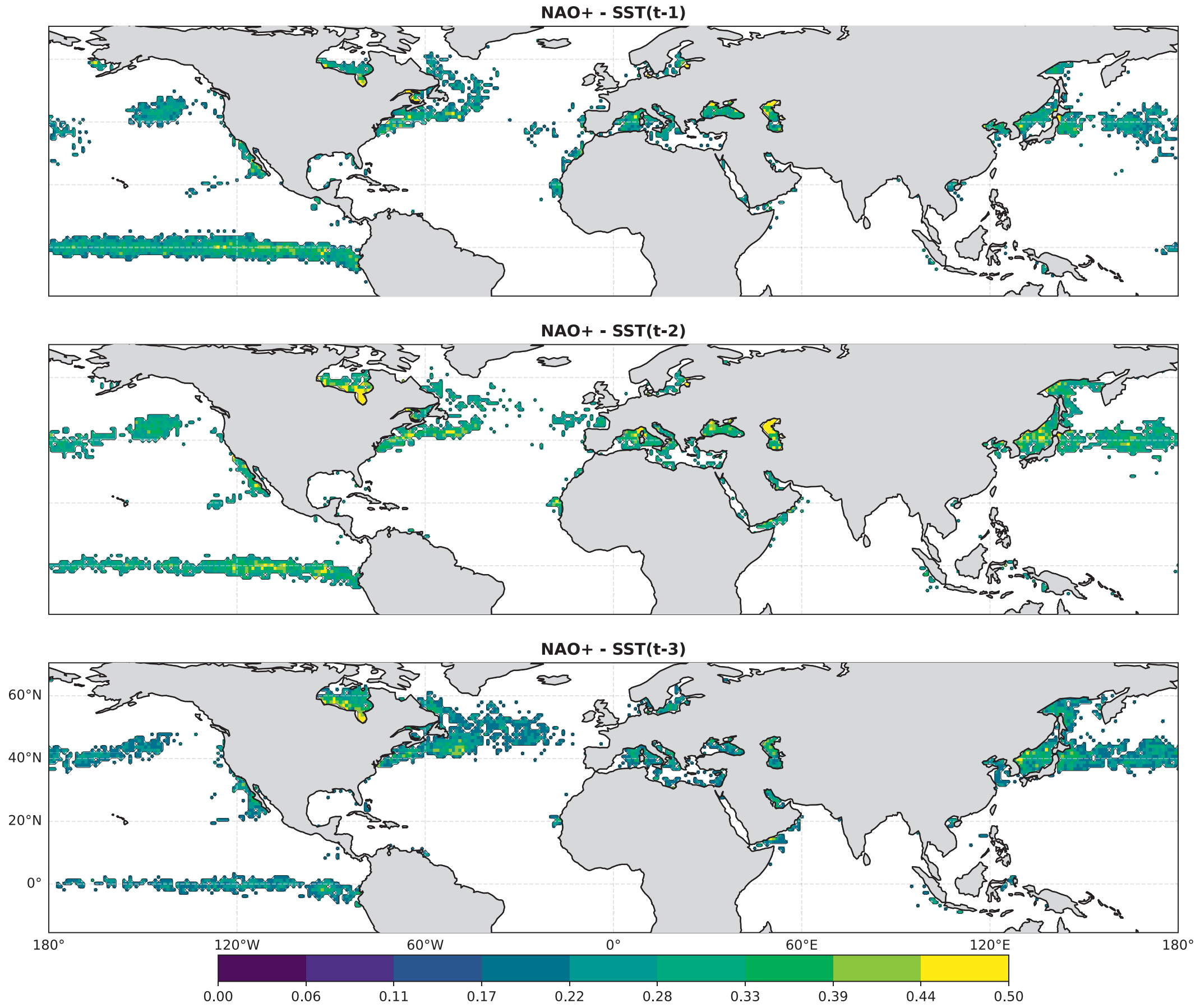

***Figure 4.*** *Average sensitivity heatmaps (unitless) for SST at monthly lags t-1, t-2, and t-3 relative to the target season t, for nine well-predicted positive NAO years (1951, 1959, 1972, 1974, 1984, 1993, 1994, 1998 and 2015). Only grid points significant at the 95% confidence level are shown in color; non-significant areas are left blank.*

Figures 4 and 5 show the average sensitivity heatmaps for SST fields 1, 2, and 3 months prior to well-predicted positive-NAO (1951, 1959, 1972, 1974, 1984, 1993, 1994, 1998 and 2015) and negative-NAO (1950, 1955, 1961, 1962, 1967, 1969, 1970, 1975 and 2001) years, respectively. A year is considered well-predicted when the absolute prediction error (|true NAO − predicted NAO|) is less than 0.6, considering only strong NAO events (|true NAO| > 0.4), leading to nine positive and negative NAO events selected for the sensitivity analysis.

In both phases, the equatorial Pacific, associated with ENSO, emerges as a key region influencing the model predictions, along with its teleconnections in the extratropical Pacific (e.g., the Kuroshio–Oyashio Extension region and the Sea of Japan). North Atlantic SSTs also play an important role, especially in the region of the North Atlantic Current and in the Hudson Bay near the sea ice edge, indicating that the model integrates signals from both tropical and extratropical regions. Despite recent results suggesting a dominant Indian Ocean contribution to the NAO (Raganato et al., 2025), the Indian Ocean plays a limited role in the CNN decision-making, with notable sensitivity only off the west coast of Sumatra and primarily for positive NAO years. For negative NAO years, the attribution is generally weaker, particularly over the tropical Pacific, likely reflecting a weaker model response.

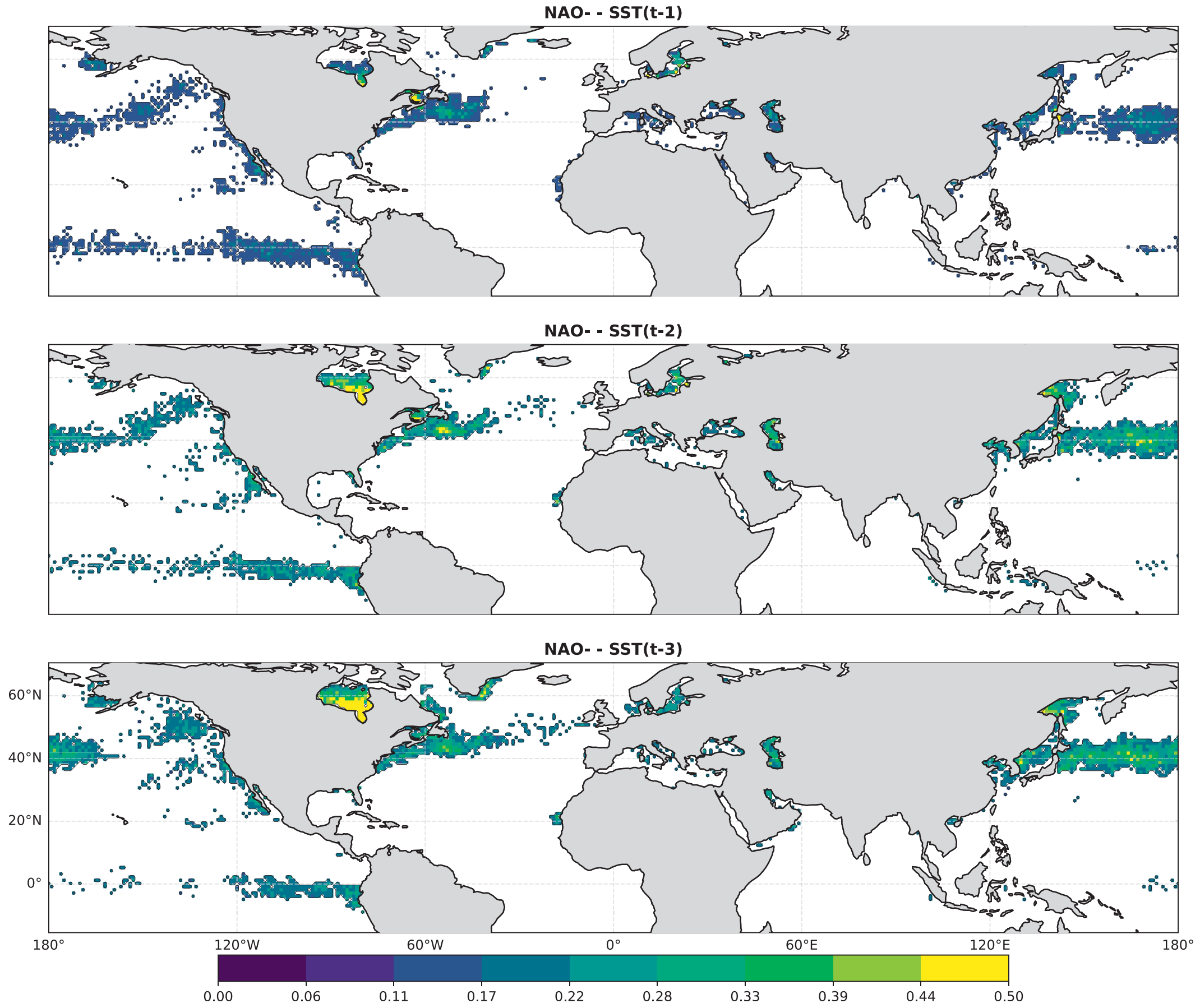

***Figure 5.*** *Average sensitivity heatmaps (unitless) for SST at monthly lags t-1, t-2, and t-3 relative to the target season t, for nine well-predicted negative NAO years (1950, 1955, 1961, 1962, 1967, 1969, 1970, 1975, and 2001). Only grid points significant at the 95% confidence level are shown in color; non-significant areas are left blank.*

### *3.2. Year-to-year variability and ENSO*

Motivated by the sensitivity analysis, which highlights the influence of ENSO on the predictions, we investigate how model performance varies across distinct ENSO conditions. ENSO intensity is quantified using the Niño3.4 index, defined as the area-weighted mean SST anomaly over 5°S-5° N, 170°W-120°W. This index is a common proxy for SST variability associated with ENSO (e.g., Deser et al., 2010). Weak/neutral years correspond to |anomaly| < 1.0 °C, while moderate/strong years correspond to |anomaly| ≥ 1.0 °C. Using this definition, we identify 22 moderate/strong years and 52 weak/neutral years. During moderate or strong ENSO years, the CNN achieves an nRMSE of 0.86 and a correlation of 0.57, while the linear model achieves lower scores of 0.91 and 0.47, respectively.

In both cases, these values are better than the average performance over the full period. In contrast, during weak or neutral ENSO years, the CNN shows an nRMSE of 0.90 and a correlation of 0.46, whereas the linear model records an nRMSE of 0.93 and a correlation of 0.43. In both cases, these values are worse than the average performance over the full period. These results indicate that ENSO conditions have an influence on model performance, though this effect is more pronounced for the CNN than for the linear model.

## 4. Conclusion

Recent advances in machine learning have shown great promise for both seasonal prediction and understanding of complex nonlinear climate processes. Previous studies, such as Kent et al. (2025), demonstrated that advanced neural network models can outperform traditional dynamical models for key indices, including the NAO. Shackelford et al. (2025) further emphasized the dominant role of ENSO-related SST variability in driving neural network prediction skill of the North Atlantic atmospheric circulation, identifying specific SST regimes linked to enhanced predictability using self-organizing maps, including both ENSO and non-ENSO sources. Building on this foundation, the present study highlights the potential of a simple neural network to predict early winter NAO at medium-range to sub-seasonal time scales by capturing complex, nonlinear relationships between SST anomalies and NAO variability. These SST-NAO links may be underestimated in climate models (Scaife et al., 2014). Using explainable AI techniques, the model's decision-making process can indicate physically relevant SST regions, providing insights into the drivers of NAO variability. The tropical Pacific Ocean appears as the one having the strongest impact on the CNN choices. Furthermore, we found that skill from the CNN depends on ENSO conditions, with higher skill observed during moderate and strong ENSO years compared to weak or neutral ENSO years. Notably, no well-predicted cases were found after 2001, with the exception of 2015, which corresponds to an exceptionally strong El Niño event. This is consistent with the positive phase of the Atlantic Multidecadal Variability (AMV) prevailing since the late 1990s, which may act as a background state modulating ENSO-NAO teleconnections (Chen et al., 2025). Whether the reduced skill in recent decades is partly related to the AMV state is an interesting avenue for future work. Furthermore, the model's tendency to underestimate the spread of NAO variability could be addressed by replacing the standard MSE loss with a quantile-based framework. Unlike MSE, which optimizes for the conditional mean and tends to shrink predictions toward the center of the distribution, quantile regression captures a broader range of outcomes by directly modelling conditional quantiles, leading to a more realistic representation of variability.

**Acknowledgments.** This work was performed using HPC resources from GENCI-IDRIS (Grant 2025-AD011013295R3).

**Author contribution.** Conceptualization: E.P., G.G.; Data curation: E.P., G.G.; Formal analysis: E.P.; Investigation: E.P., G.G., C.M., D.S., S.T.; Supervision: G.G., C.M., D.S., S.T.; Writing – original draft: E.P.; Writing – review & editing: G.G., C.M., D.S., S.T.

**Competing interests.** The authors declare none.

**Data availability statement.** Sea-Level Pressure (SLP) and Sea Surface Temperature (SST) datasets are available on the Copernicus Climate Data Store: ERA5 hourly data on single levels from 1940 to present (Hersbach et al., 2023). Codes can be found on GitLab.

**Ethics statement.** The research meets all ethical guidelines, including adherence to the legal requirements of the study country.

**Funding statement.** This work also benefited from the French state aid managed by the ANR under the "Investissements d'avenir" programme with the reference ANR-11-IDEX-0004-17-EURE-0006.

**Provenance.** This article has been developed from an abstract accepted into Climate Informatics 2025 to be reviewed according to the standard Environmental Data Science process.

## References

**Abid MA, Kucharski F, Molteni F and Almazroui M** (2023) Predictability of Indian Ocean precipitation and its North Atlantic teleconnections during early winter. *Npj Climate and Atmospheric Science 6*(1), 17. https://doi.org/10.1038/s41612-023-00328-z.

**Baehrens D, Schroeter T, Harmeling S, Kawanabe M, Hansen K and Mueller K-R** (2009) How to explain individual classification decisions. https://doi.org/10.48550/ARXIV.0912.1128.

**Barnston AG and Livezey RE** (1987) Classification, seasonality and persistence of low-frequency atmospheric circulation patterns. *Monthly Weather Review 115*(6), 1083–1126. https://doi.org/10.1175/1520-0493(1987)115<1083:CSAPOL>2.0.CO;2.

**Bommer PL, Kretschmer M, Hedström A, Bareeva D and Höhne MM-C** (2024) Finding the right xai method—A guide for the evaluation and ranking of explainable ai methods in climate science. *Artificial Intelligence for the Earth Systems 3*(3), e230074. https://doi.org/10.1175/AIES-D-23-0074.1.

**Bretherton CS**, **Widmann M**, **Dymnikov VP**, **Wallace JM and Bladé I** (1999) The effective number of spatial degrees of freedom of a time-varying field. *Journal of Climate 12*(7), 1990–2009. https://doi.org/10.1175/1520-0442(1999)012<1990:TENOSD>2.0.CO;2.

**Chen S**, **Chen W**, **Wu R**, **Yu B**, **Graf H-F**, **Cai Q**, **Ying J and Xing W** (2025) Atlantic multidecadal variability controls Arctic-ENSO connection. *Npj Climate and Atmospheric Science 8*(1), 44. https://doi.org/10.1038/s41612-025-00936-x.

**Deser C**, **Alexander MA**, **Xie S-P and Phillips AS** (2010) Sea surface temperature variability: Patterns and mechanisms. *Annual Review of Marine Science 2*(1), 115–143. https://doi.org/10.1146/annurev-marine-120408-151453.

**Deser C**, **Tomas RA and Peng S** (2007) The transient atmospheric circulation response to North Atlantic sst and sea ice anomalies. *Journal of Climate 20*(18), 4751–4767. https://doi.org/10.1175/JCLI4278.1.

**Feldstein SB** (2000) The timescale, power spectra, and climate noise properties of teleconnection patterns. *Journal of Climate 13*(24), 4430–4440. https://doi.org/10.1175/1520-0442(2000)013<4430:TTPSAC>2.0.CO;2.

**Gastineau G and Frankignoul C** (2015) Influence of the North Atlantic sst variability on the atmospheric circulation during the twentieth century. *Journal of Climate 28*(4), 1396–1416. https://doi.org/10.1175/JCLI-D-14-00424.1.

**Hardiman SC**, **Dunstone NJ**, **Scaife AA**, **Smith DM**, **Knight JR**, **Davies P**, **Claus M and Greatbatch RJ** (2020) Predictability of European winter 2019/20: Indian Ocean dipole impacts on the NAO. *Atmospheric Science Letters 21*(12), e1005. https://doi.org/10.1002/asl.1005.

**Hersbach H**, **Bell B**, **Berrisford P**, **Biavati G**, **Horányi A**, **Muñoz Sabater J**, **Nicolas J**, **Peubey C**, **Radu R**, **Rozum I**, **Schepers D**, **Simmons A**, **Soci C**, **Dee D and Thépaut J-N** (2023) ERA5 hourly data on single levels from 1940 to present. Copernicus Climate Change Service (C3S) Climate Data Store (CDS). https://doi.org/10.24381/cds.adbb2d47.

**Kent C**, **Scaife AA**, **Dunstone NJ**, **Smith D**, **Hardiman SC**, **Dunstan T and Watt-Meyer O** (2025) Skilful global seasonal predictions from a machine learning weather model trained on reanalysis data. *Npj Climate and Atmospheric Science 8*(1), 314. https://doi.org/10.1038/s41612-025-01198-3.

**King MP**, **Herceg-Bulić I**, **Bladé I**, **García-Serrano J**, **Keenlyside N**, **Kucharski F**, **Li C and Sobolowski S** (2018) Importance of late fall enso teleconnection in the euro-Atlantic sector. *Bulletin of the American Meteorological Society 99*(7), 1337–1343. https://doi.org/10.1175/BAMS-D-17-0020.1.

**Kushnir Y**, **Robinson WA**, **Bladé I**, **Hall NMJ**, **Peng S and Sutton R** (2002) Atmospheric gcm response to extratropical sst anomalies: Synthesis and evaluation*. *Journal of Climate 15*(16), 2233–2256. https://doi.org/10.1175/1520-0442(2002)015<2233:AGRTES>2.0.CO;2.

**Molteni F and Brookshaw A** (2023) Early- and late-winter ENSO teleconnections to the euro-Atlantic region in state-of-the-art seasonal forecasting systems. *Climate Dynamics 61*(5–6), 2673–2692. https://doi.org/10.1007/s00382-023-06698-7.

**Raganato A**, **Abid MA and Kucharski F** (2025) The combined link of the Indian Ocean dipole and enso with the North Atlantic-european circulation during early boreal winter in reanalysis and the ecmwf seas5 hindcast. *Journal of Climate 38*(2), 445–460. https://doi.org/10.1175/JCLI-D-23-0703.1.

**Scaife AA**, **Folland CK**, **Alexander LV**, **Moberg A and Knight JR** (2008) European climate extremes and the North Atlantic oscillation. *Journal of Climate 21*(1), 72–83. https://doi.org/10.1175/2007JCLI1631.1.

**Scaife AA**, **Arribas A**, **Blockley E**, **Brookshaw A**, **Clark RT**, **Dunstone N**, **Eade R**, **Fereday D**, **Folland CK**, **Gordon M**, **Hermanson L**, **Knight JR**, **Lea DJ**, **MacLachlan C**, **Maidens A**, **Martin M**, **Peterson AK**, **Smith D**, **Vellinga M**, **Wallace E**, **Waters J and Williams A** (2014) Skillful long-range prediction of European and north American winters. *Geophysical Research Letters 41*(7), 2514–2519. https://doi.org/10.1002/2014GL059637.

**Shackelford K**, **DeMott CA**, **Van Leeuwen PJ and Barnes EA** (2025) A regimes-based approach to identifying seasonal state-dependent prediction skill. *Journal of Geophysical Research: Atmospheres 130*(8), e2024JD042917. https://doi.org/10.1029/2024JD042917.

**Small, R. J.**, **deSzoeke, S.**, **Xie, S. P.**, **O'Neill, L.**, **Seo, H.**, **Song, Q.**, … & **Minobe, S.** (2007). Atmospheric response to ocean fronts and eddies: A Review. In *Proceedings of the 15th Conference on Air-Sea Interaction*, Portland, OR, USA (pp. 20–24).

**Wilks DS** (2006) On "field significance" and the false discovery rate. *Journal of Applied Meteorology and Climatology 45*(9), 1181–1189. https://doi.org/10.1175/JAM2404.1.

**Yuan X**, **Kaplan MR and Cane MA** (2018) The interconnected global climate system—A review of tropical-polar teleconnections. *Journal of Climate 31*(15), 5765–5792. https://doi.org/10.1175/JCLI-D-16-0637.1.

**Zhang W and Jiang F** (2023) Subseasonal variation in the winter enso-nao relationship and the modulation of tropical North Atlantic sst variability. *Climate 11*(2), 47. https://doi.org/10.3390/cli11020047.